\documentclass[twocolumn,10pt]{IEEEtran}
\usepackage{amsmath,amssymb,epsfig,graphicx}
\usepackage{theorem}
\newcommand{\be}{\begin{equation}}
\newcommand{\ee}{\end{equation}}

\title{Communication models with distributed transmission rates and buffer sizes}
\author{
David Arrowsmith$^{\dag},$\thanks{$^{\dag}$ Mathematics Research
Centre, Queen Mary, University of London, London E1 4NS, U.K.}
Mario di Bernardo$^{*\circ},$\thanks{$^{*}$Department of
Engineering Mathematics, University of Bristol, Bristol, U.K.}
Francesco Sorrentino$^{\circ \ddag}$\thanks{$^{°\circ}$Department
of Systems and Computer Science, University of Naples Federico II,
Naples, Italy} \thanks{$^{°\ddag}$Corresponding author. Email:
{\small fsorrent@unina.it}}}

\date{}

\begin{document}\maketitle
\begin{abstract}
The paper is concerned with the interplay between network
structure and traffic dynamics in a communications network, from
the viewpoint of end-to-end performance of packet transfer. We use
a model of network generation that allows the transition from
random to scale-free networks. Specifically, we are able to
consider three different topological types of networks: (a)
random; (b) scale-free with $\gamma = 3$; (c) scale-free with
$\gamma = 2$. We also use an LRD traffic generator in order to
reproduce the fractal behavior that is observed in real world data
communication. The issue is addressed of how the traffic behavior
on the network is influenced by the variable factors of the
transmission rates and queue length restrictions at the network
vertices. We show that these factors can induce drastic changes in
the throughput and delivery time of network performance and are
able to counter-balance some undesirable effects due to the
topology.
\end{abstract}

\section{Introduction}

We consider the class of networks generated by the model
introduced in \cite{korea}. This model is particularly interesting
since it allows a parametrized transition between random
(homogenous) and scale-free (heterogenous) graphs. In particular,
scale-free graphs are characterized by the presence of few very
high degree vertices, called \emph{hubs}, which are responsible
for a drastic reduction of the average distance between network
nodes.

In \cite{Ar:dB:So}, we have already studied the effects of such a
transition on the network communication performance.  The main
result shows that, somewhat surprisingly,  the structure of
scale-free graphs, which are ubiquitous in nature, does not lead
to any benefit but rather a worsening in terms of the end-to-end
performance. In particular, the characteristic parameters known as
\emph{throughput} and \emph{delivery time} were considerably
affected by the congestion at the network hubs. This is
counter-intuitive when one considers that the shortening of the
distances in the network might result in a reduction of the
delivery time and thus an increase of the throughput.

This interesting phenomenon is analogous to the \emph{paradox of
heterogeneity} \cite{paradox}, which has been observed in the
context of synchronizability of scale-free networks. Indeed, it is
unrealistic to assume that resources such as bandwidth are
uniformly distributed among the network nodes in strongly
heterogenous networks. Instead, it is very likely that hubs, which
are characterized by a high number of incoming and outgoing links,
are found to play a fundamental role in communication over the
network. They are typically characterized by having higher server
strength transmission rates and larger buffers than more
peripheral nodes. For that reason we consider here the following
variation to the study presented in \cite{Ar:dB:So}:

\begin{itemize}
    \item The transmission rate $r$ is assumed to scale with the degree at each vertex $i$, $k(i)$, as: $r(i)= c_1
    k(i)^{\alpha}$ (note that in the particular case where
    $\alpha=0$, we recover the original case, with all the nodes
     having the same transmission rates);
    \item the maximum queue length (i.e., the buffer size) is no longer assumed infinite but is
    taken to scale with the degree at each vertex $i$, $k(i)$, as: $q(i)=
    c_2 k(i)^{\beta}$.
\end{itemize}

We use the power laws above not necessarily for accurate
simulation of the nature of distributed transmission rates and
buffer sizes, but for their qualitative properties which
emphasizes the importance of these hubs. In what follows, we
analyze separately, by means of numerical simulations, the effects
of varying $\alpha$ and $\beta$ on the network communication
performance. As a representative case, we assume $c_1=1$ and
$c_2=50$. Similar behaviour was observed for other values of $c_1$
and $c_2$.

Then we compare the behaviors of networks characterized by
different topological features. More precisely, networks will be
considered with different  degree distributions; the degree $k$ at
a given node being the number of incident links at the node.
Particular emphasis will be given to scale-free topologies in
which the degree distribution is observed to follow a power law,
i.e. $P(k) \sim k^{-\gamma}$
\cite{Ba:Al99,Fa:Fa99,Am:Sc00,Do:Me02}.
\begin{figure*}
%\begin{minipage}{\textwidth}
\begin{center}
\epsfig{width=0.48\textwidth, file=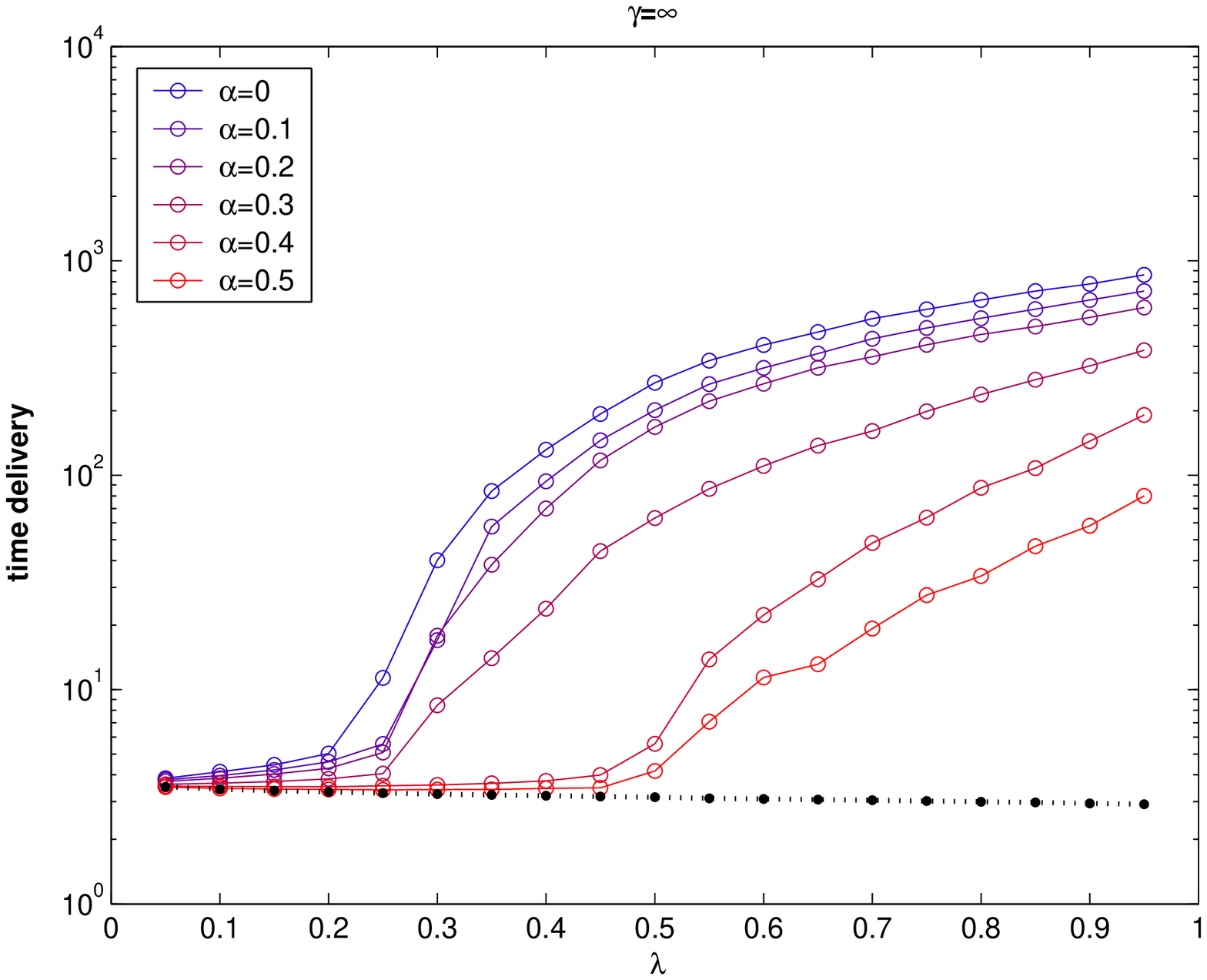}
\epsfig{width=0.48\textwidth, file=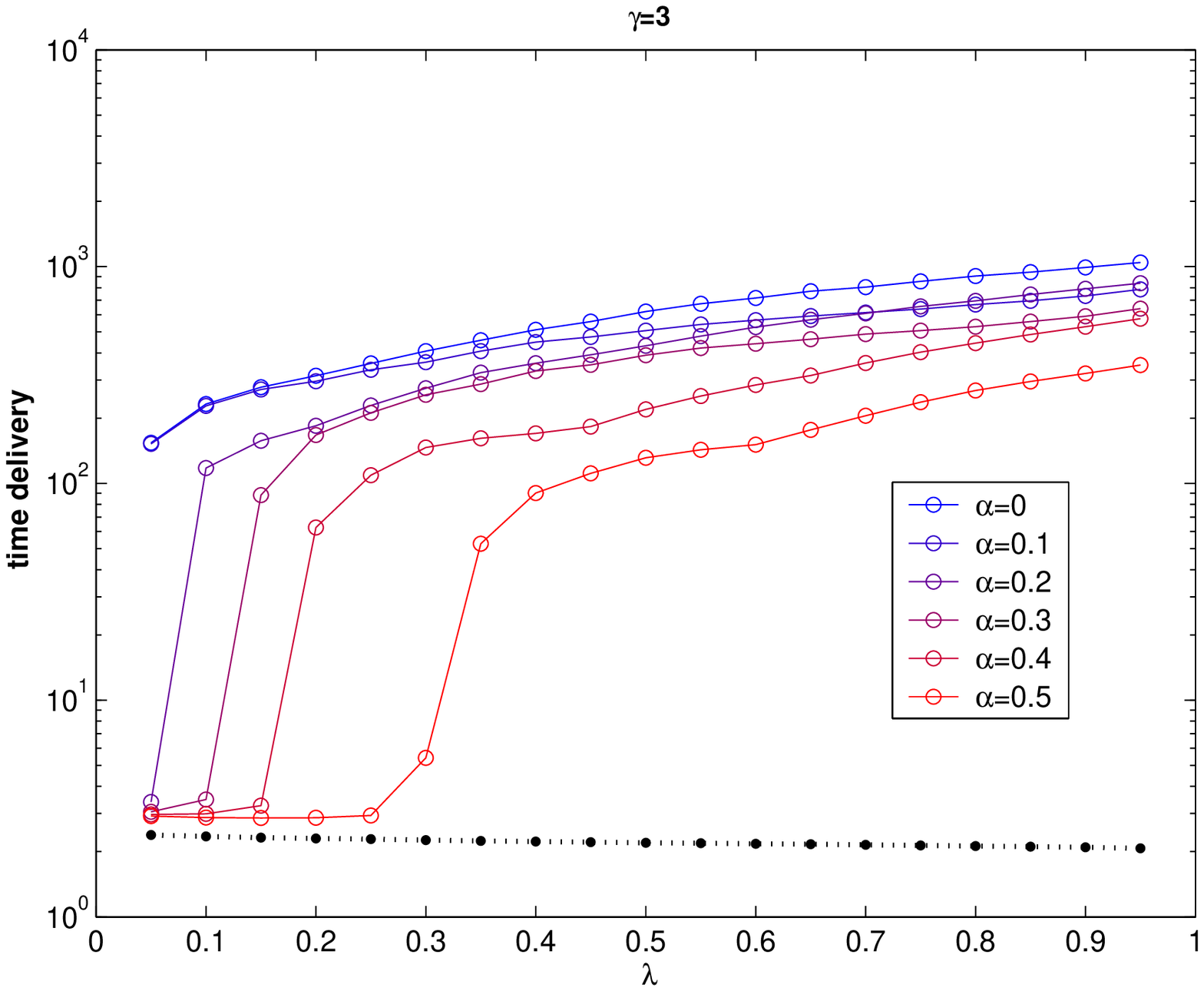}
\caption{\label{F1} \small Delivery time versus the generation
rate, $\lambda$. The network is a (a) random graph
($\gamma=\infty$), (b) scale-free graph ($\gamma=3$) with number
of nodes $N=512$ and number of edges $M=2N$. We show the effects
of varying the transmission rates $r(i)$ at node $i$, according to
the law $r(i)= k(i)^{\alpha}$, for $\alpha$ ranging between $0$
and $0.5$ (blue to red). The black dotted line represents the free
regime at $\alpha=1$.}
\end{center}
%\end{minipage}
\end{figure*}
\section{Network Topology}

We consider networks with assigned topologies overlayed by a
packet traffic communication dynamic. We try to characterize the
way in which the underlying topology can affect the network
behavior and performance by varying the values of the {\it
scale-free} degree distribution exponent $\gamma$.

There are many classical models for the production of network
topologies, but they lack the ingredient of specific degree
characteristics. The famous $Erd\ddot{o}s-Ren\acute{y}i$
\cite{Er:Re} random network algorithm leads to networks with
Poissonian degree distribution, which are characterized by an
exponential cut-off for high degrees. This is based on a very
simple procedure, namely a network with $N$ vertices and $M$ edges
is built as follows: we select with uniform probability two of the
$N$ possible vertices and link them unless they are either already
connected or self-links are generated. We repeat this iteration
$M$ times.

In order to cause the transition from random to scale-free network
we use the static model introduced in \cite{korea}. Vertices are
indexed by an integer $i$, for $(i=1....,N)$, and assigned a {\it
weight} or {\it fitness} $p_i=i^{-\eta}$ where $\eta$ is a
parameter between 0 and 1. Two different vertices are selected
with probabilities equal to the normalized weights,
$p_i/\sum_k{p_k}$ and $p_j/\sum_k{p_k}$, and an edge is added
between them unless one exists already. This process is repeated
until $M$ edges are made in the system leading to the mean degree
$\langle k \rangle = 2M/N $. This results in the expected degree
at vertex $i$ scaling as $k_i \sim (\frac{N}{i})^\eta$
\cite{korea}. We then have the degree distribution, i.e. the
probability of a vertex being of degree $k$, given by $P(k)\sim
k^{-\gamma}$ with $ \gamma= 1+ \frac{1}{\eta}$. Thus, by varying
$\eta$, we can obtain the exponent $\gamma$ in the range,
$2<\gamma<\infty$. Moreover the ER graph is generated by taking
$\eta=0$.

It is worth noting that the static model described here can be
considered as an extension of the standard ER model for generating
\emph{random} scale-free networks, i.e. networks with prescribed
degree distribution, but completely random with respect to all the
other features.

\section{Model of Network Data Traffic}

We use the family of {\it Erramilli} interval maps as the
generator for each LRD traffic source within the network
\cite{Erramilli}.  The maps are given by
$f=f_{(m_1,m_2,d)}:I\rightarrow I$, $I=[0,1]$, where:

\be \label{ESPmap} f(x)=\begin{cases}
x+(1-d)\left({x/d}\right)^{m_1}, & x\in [0,d],\cr
x-d\left((1-x)/(1-d)\right)^{m_2}, & x\in (d,1],
\end{cases}
\ee

\noindent where $d\in (0,1)$.  The map $f$ is iterated to produce
an {\it orbit}, or sequence, of real numbers $x_n\in [0,1]$ which
is then converted into a binary {\it Off-On} sequence where the
$n$-th value is  {\it `Off'} if $x_n\in [0,d]$, and {\it `On'} if
$x_n\in (d,1]$. If the orbit is in the {\it `On'} state, each
iteration of the map represents a packet generated. The parameters
$m_1,m_2\in [3/2,2]$ induce {\it map intermittency}. When
$m_1=m_2=1.5$ we have {\it short range dependent} binary output
and this becomes fully {\it long range dependent} binary output
for $m={\rm Max}\{m_1,m_2\}=2.0$. The indicator of long-range
dependence is given by the Hurst parameter $H=(3m-4)/(2m-2)$.

The network involves two types of nodes: hosts and routers. The
first are nodes that can generate and receive messages and the
second can only store and forward messages. The density of hosts,
say $\rho \in [0,1]$, is the ratio between the number of hosts and
the total number of nodes in the network (in this paper we take
$\rho = 0.16$). Hosts are randomly distributed throughout the
network.

\begin{figure*}[tbp]
%\begin{minipage}[c]{\textwidth}
\begin{center}
\epsfig{width=0.95\textwidth, file=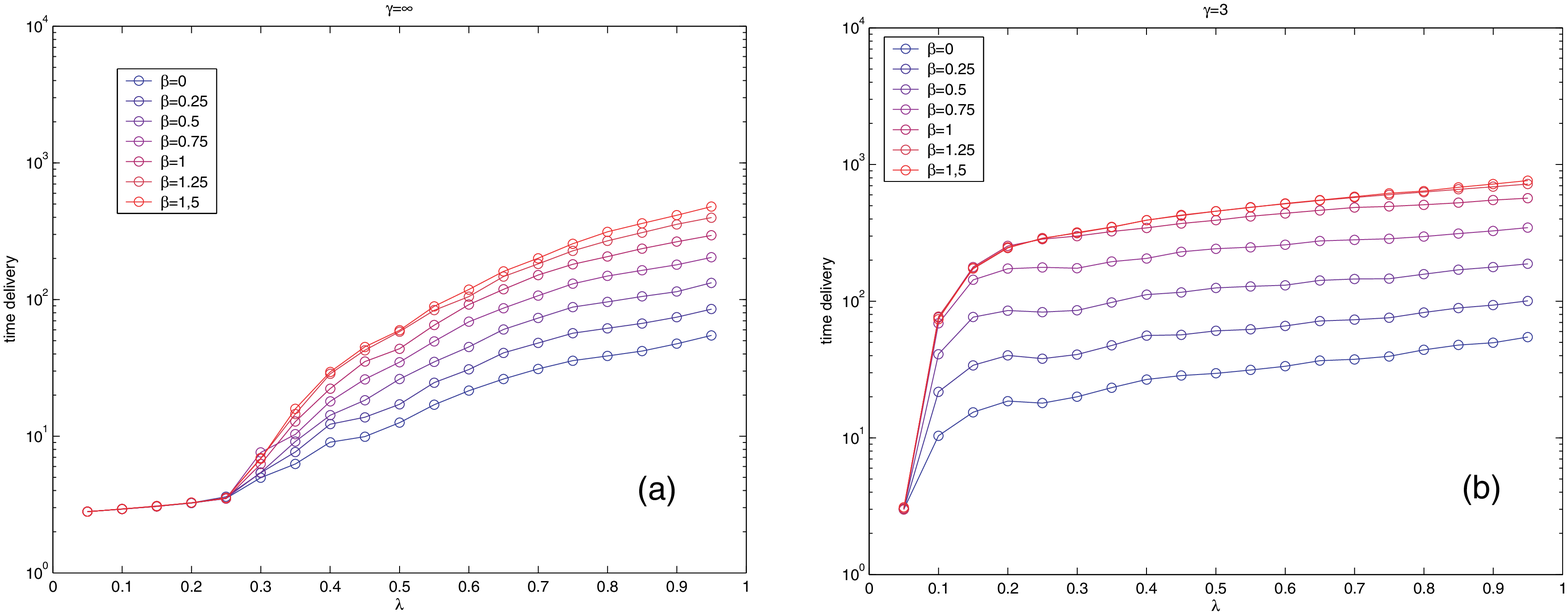}
\caption{\label{F3} \small Delivery time versus the generation
rate, $\lambda$. The network is a (a) random graph
($\gamma=\infty$), (b)  scale-free graph ($\gamma=3$) with number
of nodes $N=512$ and number of edges $M=2N$. We show the effects
of varying the queue length $q(i)$ at node $i$, according to the
law $q(i)= 50  k(i)^{\beta}$, for $\beta$ ranging between $0$ and
$1.5$ (blue to red).}
\end{center}
%\end{minipage}
\end{figure*}

\begin{figure*}
%\begin{minipage}{\textwidth}
\begin{center}
\epsfig{width=0.48\textwidth, file=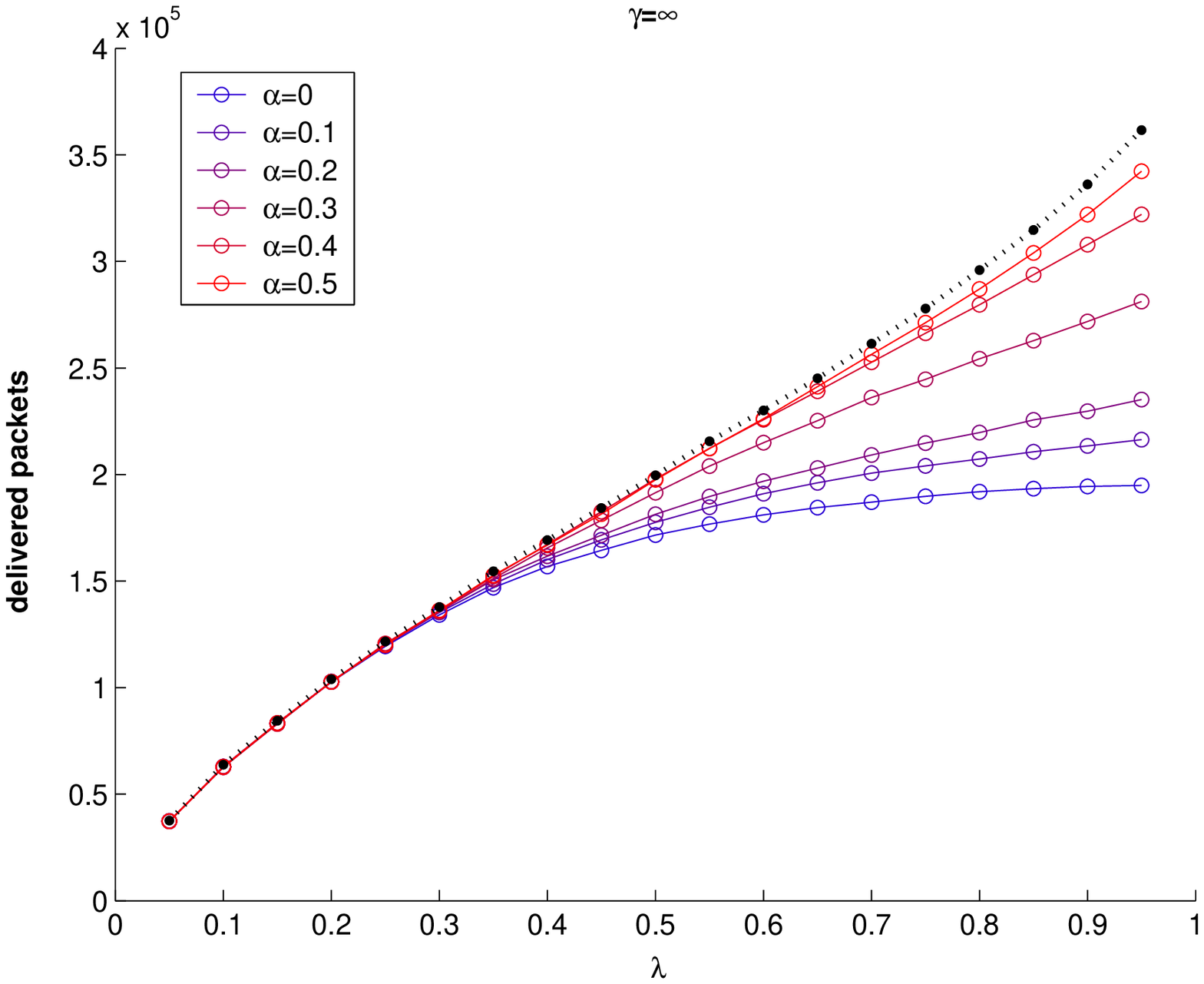}
\epsfig{width=0.48\textwidth, file=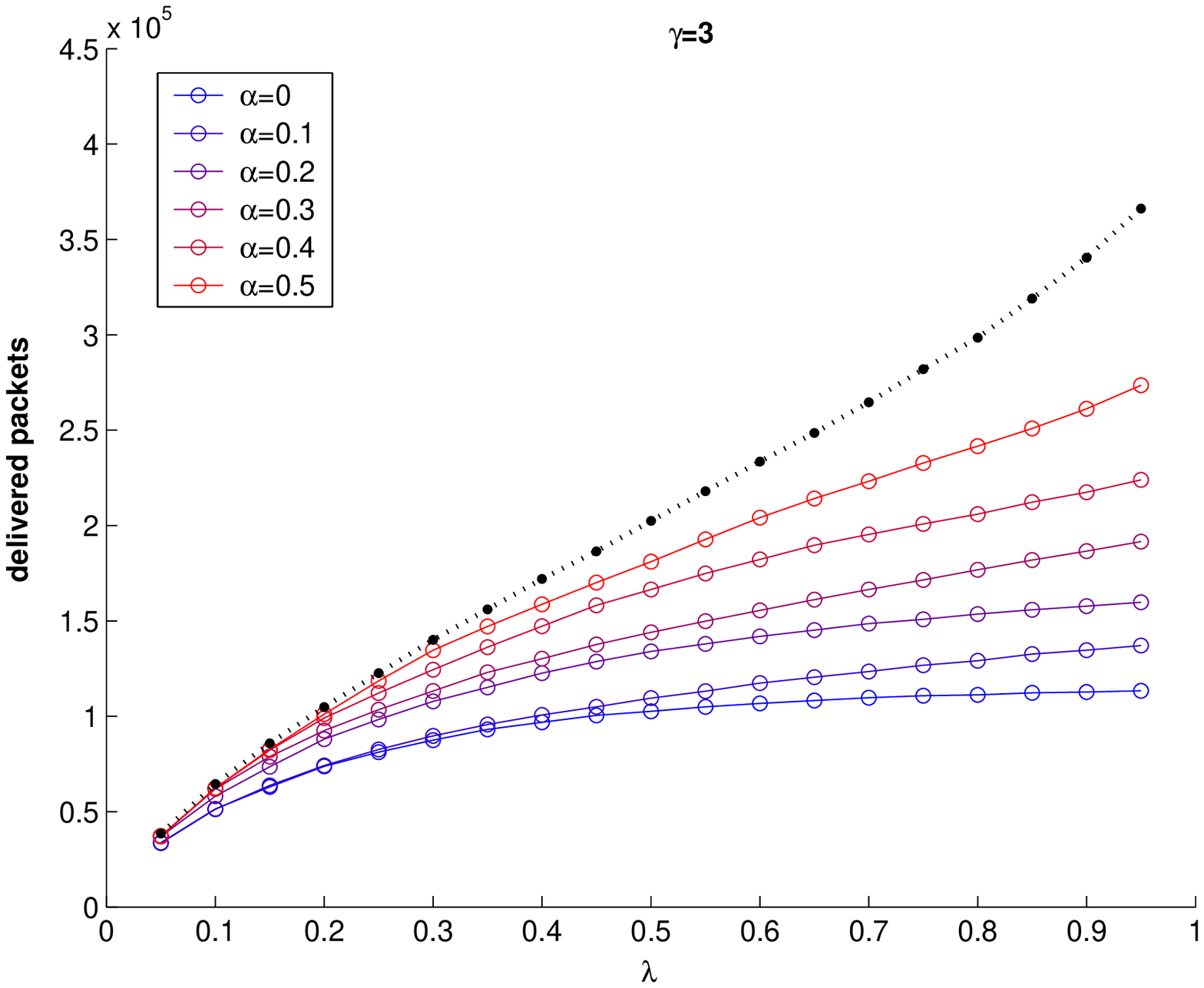}
\caption{\label{F2} \small  Number of delivered packets versus the
generation rate, $\lambda$. The network is a (a) random graph
($\gamma=\infty$), (b)  scale-free graph ($\gamma=3$) with number
of nodes $N=512$ and number of edges $M=2N$. We show the effects
of varying the transmission rates $r(i)$ at node $i$, according to
the law $r(i)= k(i)^{\alpha}$, for $\alpha$ ranging between $0$
and $0.5$ (blue to red). The black dotted line represents the free
regime at $\alpha=1$.}
\end{center}
%\end{minipage}
\end{figure*}

Packets enter the queue from one side (the {\it end}) and leave it
from the other one (the {\it head}). The head of the queue
contains a variable number of packets equal to the transmission
rate at the node. The queue at each node $i$, has a finite maximum
length, $q(i)$. If the packets arriving at the queue at vertex $i$
result in the number of packets exceeding $q(i)$, then the excess
packets are dropped.
%(?from the head or from the end, in the
%simulations I've considered from the head, probably the simulations involving the queue limits should be redone.)
%entering the queue were more than than the maxim

A {\it routing} algorithm is needed to model the dynamic aspects
of the network. Packets are created at hosts and sent through the
network one step at a time until they reach their destination
host.

The routing algorithm operates as follows:

(1) First, a host creates a packet following a distribution
defined by a chaotic map (LRD), as described above. If a packet is
generated it is put at the end of the queue for that host. This is
repeated for each host in the lattice.

(2) Packets at the head of each queue are picked up and sent to a
neighboring node selected according to the following rules. (a) A
neighbor closest to the destination node is selected. (b) If more
than one neighbor is at the minimum distance from the destination,
the link through which the smallest number of packets have been
forwarded is selected. (c) If more than one of these links shares
the same minimum number of packets forwarded, then a random
selection is made.

(3) Packets at the head of each queue, exceeding its maximum
capability, are dropped.

This process is repeated for each node in the graph. The whole
procedure of packet generation and movement represents one time
step of the simulation.

\begin{figure*}[tbp]
%\begin{minipage}{\textwidth}
\begin{center}
\epsfig{width=0.92\textwidth, file=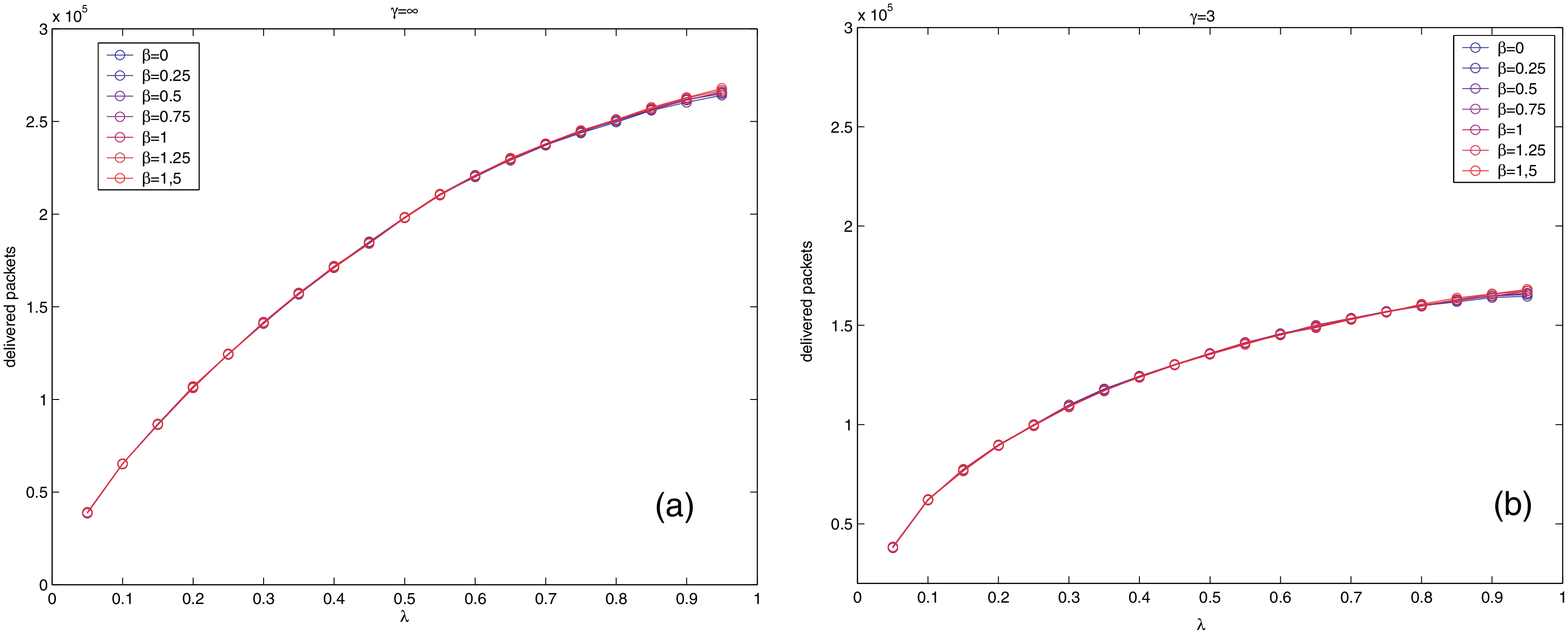}
\caption{\label{F4} \small Number of delivered packets versus the
generation rate, $\lambda$. The network is a (a) random graph
($\gamma=\infty$), (b)  scale-free graph ($\gamma=3$) with number
of nodes $N=512$ and number of edges $M=2N$. We show the effects
of varying the queue length $q(i)$ at node $i$, according to the
law $q(i)= 50  k(i)^{\beta}$, for $\beta$ ranging between $0$ and
$1.5$ (blue to red).}
\end{center}
%\end{minipage}
\end{figure*}

\section{Network Performance}
Using the network model and traffic generator detailed above,
simulations were carried out to analyse various aspects of
end-to-end performance for two types of network. Namely, results
for random graphs have been paired with those of scale-free graphs
with $\gamma=3$. We have calculated the corresponding output for
scale-free graphs with $\gamma=2$ and have found that the
differences in behaviour with the alternative value $\gamma=3$ are
negligible by comparison with the behaviour of the random graph,
and so the third set of comparisons is not repeated here.

In Fig. 1 we see that random graphs respond  more quickly with
 smaller delivery times as $\alpha$ increases (from zero).
Fig. 2 shows that the communication is much more efficient in
terms of delivered packets at high loads (or generation rate) as
$\alpha$ increases. Moreover, as shown in Fig. 3 (where the
effects of variable transmission rates have been highlighted),
 scale-free networks behave worse than random graphs for a
sufficiently high value of the parameter $\alpha$ (of the order of
unity, from our simulations). The number of delivered packets,
shown in Fig. 4, instead, is observed to be unaffected by the
buffer sizes at the nodes, being mainly determined by the network
topology. Finally, in Fig. 5, the number of dropped packets is
observed to decrease as the buffer sizes are scaled more
sensitively with vertex degree.

\section{Conclusions}
We have analyzed several features of a representative model of
communication  networks contrasting the network performance when
the graph considered is random or scale-free. The results show
that there is no obvious benefit for communication networks having
scale-free growth patterns across all performance indicators.
Nevertheless, the graphs reported here show several critical
phenomena with a set of threshold values whose analytical
investigation will be the subject of future work.

\begin{figure}[t]
\begin{center}
\epsfig{width=0.42\textwidth, file=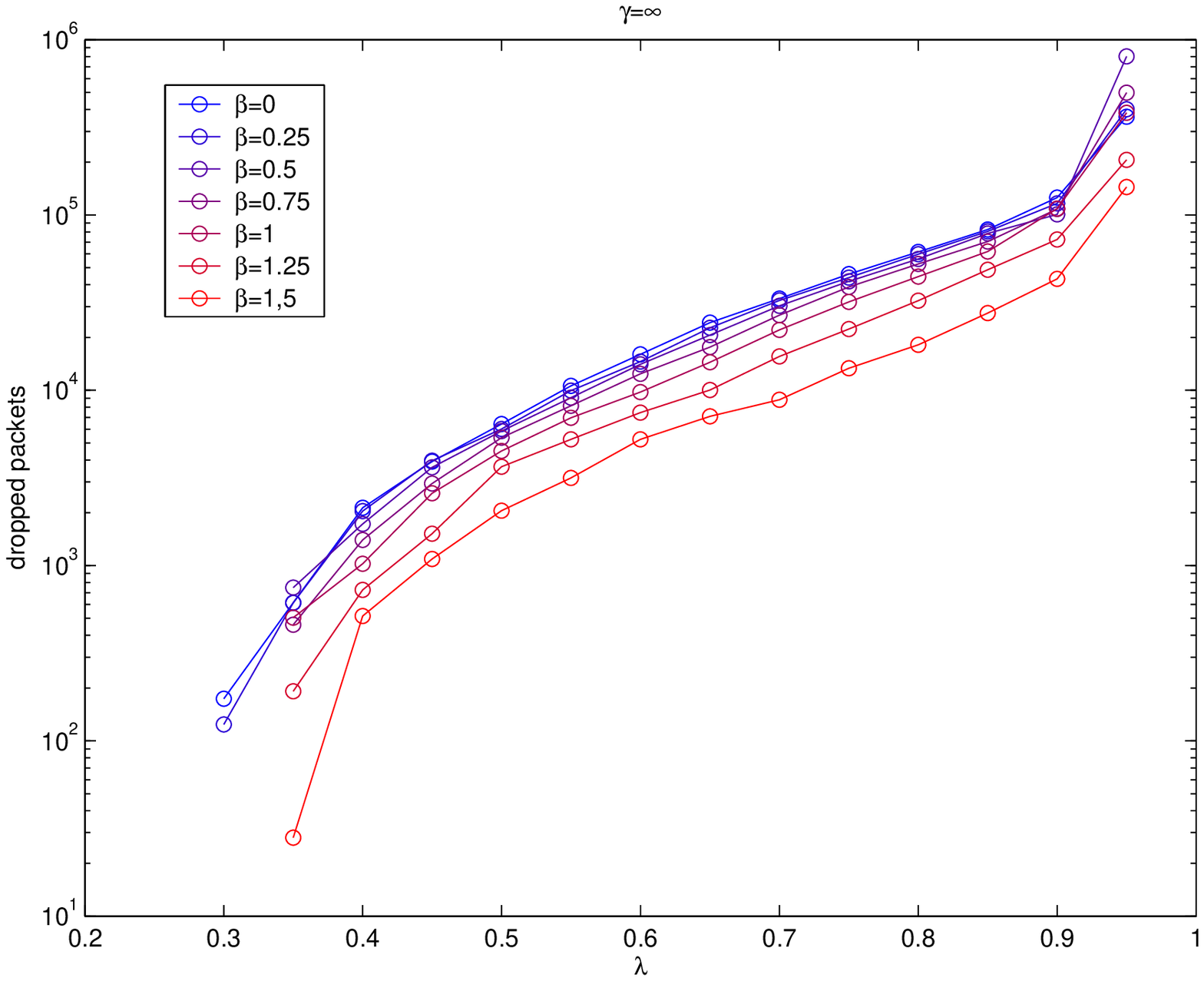} \epsfig{width=0.42
\textwidth, file=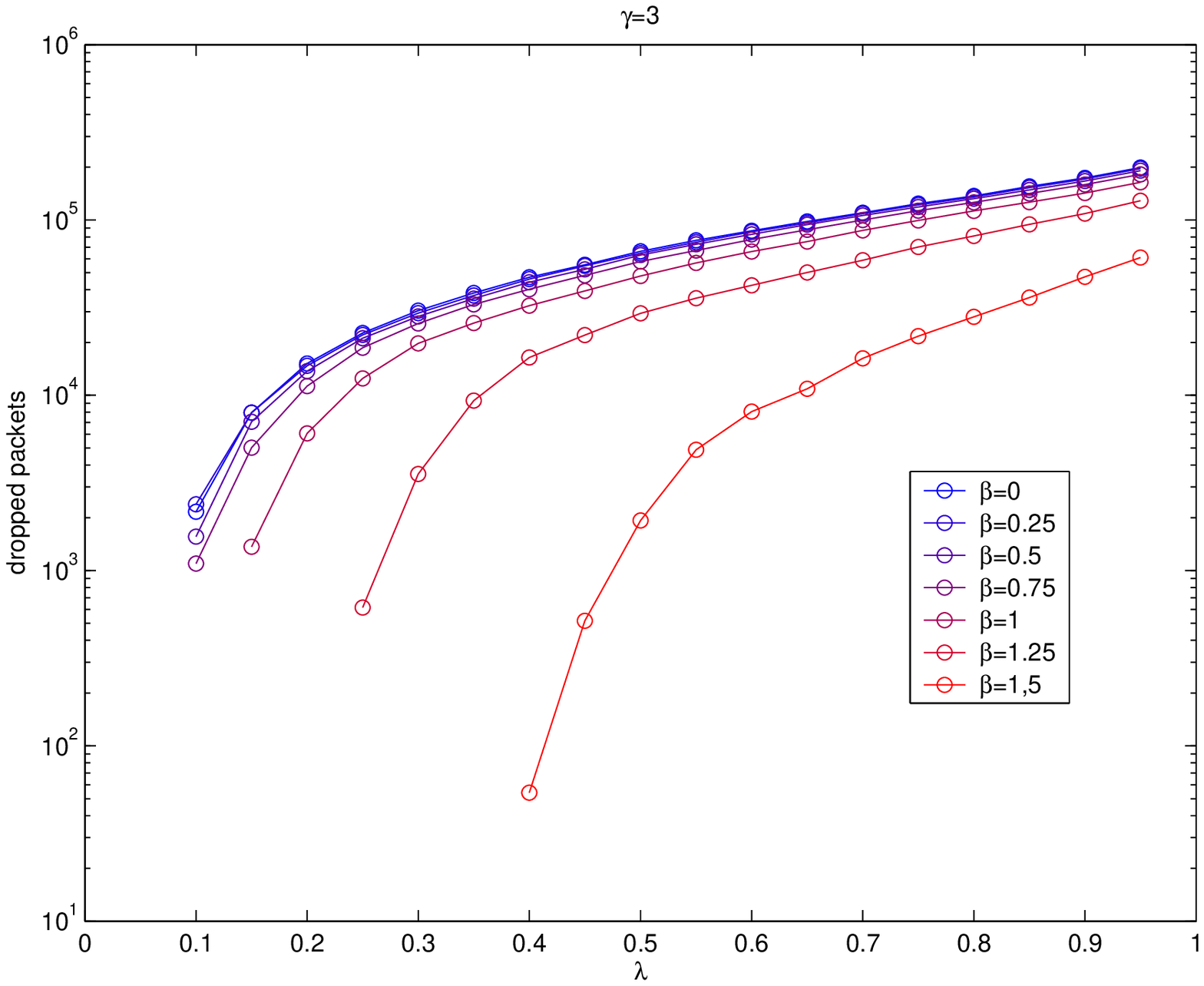}
\end{center}
\caption{\label{F14} \small Number of dropped packets versus the
generation rate, $\lambda$. The considered network is scale-free
with $\gamma=3$ and $\gamma=\infty$, the number of nodes being
$N=512$ and number of edges $M=2N$. We show the effects of varying
the queue length $q(i)$ at node $i$, according to the law $q(i)=
50 k(i)^{\beta}$, for $\beta$ ranging between $0$ and $1.5$.}
\end{figure}

%\begin{figure*}[tbp]
%%\begin{minipage}{1\textwidth}
%\begin{center}
%\epsfig{width=\textwidth, file=SIMERee2}
%%\epsfig{width=0.47\textwidth, file=fileart4}
%\caption{\label{F5} \small Number of dropped packets versus the
%generation rate, $\lambda$. The network is a (a) random graph
%($\gamma=\infty$), (b) scale-free graph ($\gamma=3$) with number
%of nodes $N=512$ and number of edges $M=2N$. We show the effects
%of varying the queue length $q(i)$ at node $i$, according to the
%law $q(i)= 50  k(i)^{\beta}$, for $\beta$ ranging between $0$ and
%$1.5$ (blue to red).}
%\end{center}
%%\end{minipage}
%\end{figure*}


\begin{thebibliography}{1}

\bibitem{korea}
K.-I. Goh, B.~Kahng, and D.Kim,
\newblock ``Universal behavior of load distribution in scale-free networks,''
\newblock {\em Phys.Rev.Lett.}, vol. 87, no. 27, 2001.

\bibitem{Ar:dB:So}
D.K. Arrowsmith, M.~di~Bernardo, and F.Sorrentino,
\newblock ``Effects of variation of load distribution on network performance,''
\newblock {\em Proc. IEEE ISCAS, Kobe, Japan}, 2005.

\bibitem{paradox}
A.E. Motter, C.~Zhou, and J.Kurths,
\newblock ``Network synchronization, diffusion, and the paradox of
  heterogeneity,''
\newblock {\em Phys.Rev.E}, vol. 71, no. 016116, 2005.

\bibitem{Ba:Al99}
A.L.Barabasi and R.Albert,
\newblock ``Emergence of scaling in random networks,''
\newblock {\em Science}, vol. 286, pp. 509--512, 1999.

\bibitem{Fa:Fa99}
M.Faloutsos, P.Faloutsos, and C.Faloutsos,
\newblock ``What does internet look like?,''
\newblock {\em Comput.Commun.Rev.}, vol. 29, pp. 251--263, 1999.

\bibitem{Am:Sc00}
L.A.N. Amaral, A.~Scala, M.Berthèlemy, and H.E.Stanley,
\newblock ``Classes of small-world networks,''
\newblock {\em Proc.Natl.Acad.Sci.USA}, vol. 97, pp. 11149--11152, 2000.

\bibitem{Do:Me02}
S.N.Dorogovstev and J.F.F. Mendes,
\newblock ``Evolution of networks,''
\newblock {\em Adv. Phys}, vol. 1079, no. 51, pp. 1079--1187, 2002.

\bibitem{Er:Re}
P.~Erdos and A.Renyi,
\newblock ,''
\newblock {\em Publ. Math. Inst. Hung. Acad.}, vol. 5, no. 17, 1960.

\bibitem{Erramilli}
A.~Erramilli, R.P. Singh, and P.~Pruthi,
\newblock ``Chaotic maps as models of packet traffic,''
\newblock {\em Proc. $14^{th}$ Int. Teletraffic Conf.}, 1994,
\newblock North--Holland (Elsevier).

\end{thebibliography}
\end{document}